\documentclass[twocolumn,trackchanges]{aastex63}
\usepackage{CJK}
\usepackage{graphicx}
\usepackage{lineno}
\graphicspath{{figures/}}
\usepackage{amsmath}
\usepackage{amssymb}
\usepackage{float}
\usepackage{color}
\usepackage{hyperref}

\usepackage{acro}

\DeclareAcronym{GR}{
	short = GR,
	long  = general relativity
	}
\DeclareAcronym{BH}{
	short = BH ,
	long  = black hole
}
\DeclareAcronym{BBH}{
	short = BBH ,
	long  = binary black hole
}
\DeclareAcronym{BNS}{
	short = BNS ,
	long  = binary neutron star
}

\DeclareAcronym{GW}{
	short = GW ,
	long  = gravitational wave
}

\DeclareAcronym{CBC}{
	short = CBC,
	long  = compact binary coalescence
}

\DeclareAcronym{PSD}{
	short = PSD,
	long  = power spectral density
}

\usepackage[capitalise]{cleveref}
\crefname{figure}{Fig.}{Figs.}
\Crefname{figure}{Fig.}{Figs.}

\def\be{\begin{equation}}
\def\ee{\end{equation}}
\def\bn{\begin{eqnarray}}
\def\en{\end{eqnarray}}
\def\({\left(}
\def\){\right)}
\def\[{\left[}
\def\]{\right]}

\newcommand{\bqn}{\begin{eqnarray}}
\newcommand{\eqn}{\end{eqnarray}}

\newcommand{\f}{\frac}
\newcommand{\p}{\partial}

\newcommand{\msun}{M_\odot}

\begin{document}
\begin{CJK*}{UTF8}{gbsn}
\title{Prospects for Detecting Gravitational Waves from Eccentric Subsolar Mass Compact Binaries}

\correspondingauthor{Yi-Fan Wang (王一帆)}
\email{yifan.wang@aei.mpg.de}

\author[0000-0002-2928-2916]{Yi-Fan Wang (王一帆)}
\author[0000-0002-1850-4587]{Alexander H. Nitz}
\affil{Max-Planck-Institut f{\"u}r Gravitationsphysik (Albert-Einstein-Institut), D-30167 Hannover, Germany}
\affil{Leibniz Universit{\"a}t Hannover, D-30167 Hannover, Germany}
\begin{abstract}
Due to their small mass, subsolar mass black hole binaries would have to be primordial in origin instead of the result of stellar evolution.
Soon after formation in the early universe, primordial black holes can form binaries after decoupling from the cosmic expansion.
Alternatively, primordial black holes as dark matter could also form binaries in the late universe due to dynamical encounters and gravitational-wave braking.
A significant feature for this channel is the possibility that some sources retain nonzero eccentricity in the LIGO/Virgo band.
Assuming all dark matter is primordial black holes with a delta function mass distribution, $1\msun-1\msun$ binaries formed in this late-universe channel can be detected by Advanced LIGO and Virgo with their design sensitivities at a rate of $\mathcal{O}(1)$/year, where $12\%(3\%)$ of events have eccentricity at a gravitational-wave frequency of 10 Hz, $e^\mathrm{10Hz}\geq0.01(0.1)$, and nondetection can constrain the binary formation rate within this model.
Third generation detectors would be expected to detect subsolar mass eccentric binaries as light as $0.01 \msun$ within this channel, if they accounted for the majority of the dark matter.
Furthermore, we use simulated gravitational-wave data to study the ability to search for eccentric gravitational-wave signals using a quasi-circular waveform template bank with Advanced LIGO design sensitivity.
For a match-filtering targeted search, assuming binaries with a delta function mass of $0.1(1)\msun$ and the eccentricity distribution derived from this late-universe formation channel,  $41\%(6\%)$ of the signals would be missed compared to the ideal detection rate due to the mismatch in the gravitational-wave signal from eccentricity.

\end{abstract}

\section{Introduction} \label{sec:intro}
Gravitational-wave astronomy has gradually evolved into a routine method for observing compact binaries comprising black holes and/or neutron stars.
Up to now, over 50 gravitational-wave events have been identified by Advanced LIGO~\citep{TheLIGOScientific:2014jea} and Virgo~\citep{TheVirgo:2014hva}, all from compact binary coalescences \citep{Nitz:2019hdf,2010.14527}.
These detections have had significant astrophysical implications, including inferring stellar population properties \citep{2010.14533}, testing the validity of general relativity \citep{2010.14529}, and determining the value of the Hubble constant \citep{1710.05835}.

In addition to routine detection, gravitational-wave data analysis is also searching for exotic hypothesized objects, such as primordial black holes \citep{1808.04771,1904.08976,Nitz:2020}. 
Primordial black holes are hypothesized to form by direct collapse in dense regions of the early universe \citep{ 1967SvA....10..602Z,Hawking:1971,Carr:1974} and are consistent with the properties required for a candidate of cold dark matter~\citep{Carr:2020xqk}.
A variety of astrophysical observations including gravitational lensing have searched for primordial black holes and placed upper limits on the abundance with their null results, which shows that primordial black holes with a single mass are not likely to account for all dark matter (see, e.g., \cite{Carr:2020xqk} and \cite{2007.10722} for recent reviews). 
Nevertheless, as gravitational-wave astronomy provides new opportunities to observe stellar-mass black holes, we investigate the event rate and prospects for searching for gravitational waves emitted from binary primordial black holes in the LIGO/Virgo sensitive band ($\sim10-1000$ Hz). In particular, we focus on subsolar mass binary mergers with nonzero eccentricity.

There are two viable ways for primordial black holes to form binaries.  
In the early universe, after their initial formation, a nearby pair of primordial black holes can form a binary if the gravitational attraction is strong enough to decouple them from the cosmic expansion \citep{nakamura:1997,sasaki:2016,Boehm:2020jwd}. 
To be detected by gravitational-wave detectors, the binaries would have had sufficient time to circularize their orbits, thus the gravitational-wave signals are not expected to have eccentricity by the time they enter the LIGO and Virgo sensitive band. 
On the other hand, primordial black hole binaries may also form by dynamical capture due to gravitational-wave braking in the late-universe in dark matter halos.
A significant feature for this formation channel is the retention of nonzero eccentricity in the LIGO/Virgo band if the binaries form at close separation. This latter scenario is the focus of this work.

\cite{Bird:2016} has studied the event rate of GW150914-like, i.e., a $30 \msun$ binary black hole merger.
Their results show that the event rate is consistent with the empirical rate estimated from GW150914 \citep{gw150914rate,gw150914rate2}. 
However, it is challenging to confirm that GW150914 is indeed primordial in origin and not the result of stellar evolution, leaving the question still open.
\cite{Cholis:2016} has investigated the eccentricity distribution for $30\msun$ binaries in the late-universe scenario.
Meanwhile, given conventional stellar evolution models, subsolar mass ($\le 1 \msun$) compact binary coalescence can only be due to primordial black holes, instead of stellar products, because they are lighter than the minimum mass of neutron stars \citep{minmass1,minmass2}.
Therefore, in this work we only consider the mass range $[0.01,1]~\msun$, which is a smoking gun for primordial black holes, extending the work of \cite{Bird:2016} and \cite{Cholis:2016} to subsolar mass region.
Primordial black holes in this mass range are predicted by a variety of formation theories, e.g., early-universe Quantum Chromodynamics (QCD) phase transition \citep{Jedamzik:1996mr,Byrnes:2018clq,Carr:2019kxo}.

In Section \ref{sec:model}, assuming a delta function distribution of mass, we first estimate the event rate and eccentricity distribution of subsolar mass primordial black hole binaries formed in dark matter halos . The merger rate of subsolar mass binaries is shown to be higher than that of $30\msun$ binary primordial black holes as considered in \cite{Bird:2016} by one to two orders of magnitude. 
With design sensitivity, the Advanced LIGO and Virgo observatories would be expected to detect one $1\msun-1\msun$ binary with eccentricity  $e^\mathrm{10Hz}\ge0.1$ at a gravitational-wave frequency of 10 Hz with $\sim10$ years of observation, if the fraction of the primordial black hole in dark matter is $100\%$.
Third generation detectors would detect primordial black hole binaries with a single mass varying from $0.01\msun$ to $1\msun$ and eccentricity distribution derived from this late-universe formation channel.
Nondetection in the future would put constraints on the event rate modeling. 
In Section \ref{sec:search}, we investigate the capability of the current match-filtering search pipeline using a quasi-circular waveform template bank to identify signals with eccentricity by simulated gravitational-wave data.
Results show that $41\%$ and $6\%$ of the signals would be missed compared to the idealized maximum for $0.1\msun-0.1\msun$ and $1\msun-1\msun$ binaries, respectively, arising from the mismatch of gravitational-wave signals from eccentricity.
Section \ref{sec:con} presents the discussions and conclusions.

\section{Event rate and eccentricity distribution}\label{sec:model}

In this section we consider the event rate and eccentricity distribution of compact binaries formed in the late universe through two-body encounters for a delta function mass distribution which we allow to vary within $[0.01,1]\msun$.
More complex mechanisms may also induce eccentric binaries, such as three-body interaction via Kozai-Lidov effects \citep{Kozai,Lidov}, but they are not considered in this work and investigations will be left for the future.

We briefly review the two-body dynamics for primordial black hole binary formation following \cite{maggiore2008gravitational,Peters:1963,Turner:1977,Leary:2009,Bird:2016,Cholis:2016}.
Specifically, we derive the overall event rate for primordial black hole binary coalescence with a component mass $M_\mathrm{PBH}=$  $0.01/0.1/1~\msun$. To aid in comparison to other works in this field, we state our eccentricity distributions at a fiducial dominant-mode gravitational-wave frequency of $10$ Hz, which is denoted by $e^\mathrm{10Hz}$.

Two black holes can form a binary in an encounter due to gravitational-wave emission. The binary formation criterion is that the released gravitational-wave energy $\delta E_\mathrm{GW}$ exceeds the two-body kinetic energy
\be\label{eq:bcondition}
\delta E_\mathrm{GW} \geq \frac{1}{2} \mu v_\mathrm{rel}^2,
\ee
where $\mu = m_1m_2/(m_1+m_2)$ is the reduced mass, $m_1$ and $m_2$ are the component masses of the binary, and $v_\mathrm{rel}$ is the relative velocity of two black holes.
In this work we only consider equal-mass primordial black hole binaries, therefore we set $m_1=m_2=M_\mathrm{PBH}$.

At a Newtonian order, the released gravitational-wave energy is \citep{Peters:1963,Turner:1977}
\be\label{eq:deltaE}
\delta E_\mathrm{GW} = \frac{8}{15} \frac{G^{7/2}}{c^5}\frac{(m_1+m_2)^{1/2}m_1^2m_2^2}{r_p^{7/2}}f(e)
\ee
where $G$ is the Newton gravitational constant, $c$ is the speed of light, $r_p$ is the closest distance at encounter, which is the pericenter distance for an elliptical orbit, and $f(e)$ is a function of eccentricity $e$.
By inserting $f(e) = 425\pi/(32\sqrt{2})$ when $e=1$ into \cref{eq:bcondition,eq:deltaE}, one obtains the cross section for binary formation
\be\label{eq:cs}
\sigma = 2\pi \left(\frac{85\pi}{6\sqrt{2}}\right)^{2/7}\frac{G^2(m_1+m_2)^{10/7}(m_1m_2)^{2/7}}{c^{10/7}v_\mathrm{rel}^{18/7}}
\ee
where $\sigma = \pi b_\mathrm{max}^2$ by definition, and $b_\mathrm{max}$ is the maximum impact parameter.
Meanwhile, the impact parameter $b$ is related to $r_p$ by
\be\label{eq:rp}
r_p = \frac{b^2 v_\mathrm{rel}^2}{2G(m_1+m_2)}
\ee
when $G(m_1+m_2)\gg b v_\mathrm{rel}^2$, which is satisfied in this study.

Since we aim to investigate the scenario where primordial black holes are dark matter, we assume the number density of primordial black holes follows that of dark matter.
Specifically, for a dark matter halo with virial mass $M_\mathrm{vir}$, we use the Navarro-Frenk-White (NFW) density profile  \citep{Navarro:Frenk:White:1996} to model the mass density
\be
\rho_\mathrm{NFW}(r) = \frac{\rho_c}{\frac{r}{r_c}\(1+\frac{r}{r_c}\)^2},
\ee
where $r$ is the radial distance to the halo center, and $\rho_c$ and $r_c$ are the characteristic mass density and characteristic radius of the NFW profile, respectively.

The binary primordial black hole formation rate is given by \cite{Bird:2016} as
\be\label{eq:rhalo}
\mathcal{R}(M_\mathrm{vir}) = 4\pi\int_0^{R_\mathrm{vir}}\frac{r^2}{2}\( \frac{f_\mathrm{PBH}\rho_\mathrm{NFW}(r)}{M_\mathrm{PBH}}\)^2\langle\sigma v_\mathrm{rel}\rangle dr,
\ee
where $R_\mathrm{vir}$ is the halo virial radius, $f_\mathrm{PBH}$ is the mass fraction of primordial black holes in dark matter, and the angle bracket is averaged with respect to the relative velocity distribution, which is modeled to be Maxwell-Boltzmann distribution
\be\label{eq:pvrel}
P(v_\mathrm{rel}) = N_0 \[ \exp{\(-\frac{v_\mathrm{rel}^2}{v_\mathrm{max}^2}\)} - \exp{\(-\frac{v_\mathrm{vir}^2}{v_\mathrm{max}^2}\)}\]
\ee
in which $v_\mathrm{vir}$ is the circular velocity at virial radius, $v_\mathrm{max}$ is the maximum circular velocity, and $N_0$ is the normalization factor.
As demonstrated by \cite{Bird:2016}, the merger happens shortly after the formation of a binary, therefore the formation rate (\cref{eq:rhalo}) is identical to the event rate for coalescence of binary primordial black holes. 
\cref{eq:rhalo} can be integrated analytically, the result is 
\begin{eqnarray}
\mathcal{R}(M_\mathrm{vir}) &=& \(
\frac{85\pi}{6\sqrt{2}}
\)^{2/7}
\frac{2\pi G^2 f_\mathrm{PBH}^2M_\mathrm{vir}^2 D(v_\mathrm{max})}{3c r_s^3 g^2(C)}\\ \nonumber
&\times&\[ 1 - \frac{1}{(1+C)^3}\],
\end{eqnarray}
where $C=R_\mathrm{vir}/r_c$ is the concentration parameter for dark matter halo, $g(C) = \ln (1+C) - C/(1+C)$ and 
\be
D(v_\mathrm{rel}) = \int_0^{v_\mathrm{vir}} P(v_\mathrm{rel};v_\mathrm{max})
\(\frac{2v}{c}\)^{3/7} dv.
\ee

Finally, the overall event rate for binary primordial black hole coalescence is obtained by taking the sum of the contributions from all dark matter halos
\be\label{eq:rpbh}
\mathcal{R}_\mathrm{PBH} = \int 
\mathcal{R}(M_\mathrm{vir})  \frac{dN}{dM_\mathrm{vir}} dM_\mathrm{vir}
\ee

We have made use of the python code \texttt{colossus} \citep{Diemer:2018} to generate the dark matter halo mass function $dN/dM_\mathrm{vir}$ using the parameterized  formula fitting to an N-body simulation proposed by \cite{Tinker:2008}, and the concentration-dark matter halo mass relation proposed by \cite{Prada:2012}.
We have tested systematics by using  \cite{Reed07} for N-body simulation fits for the dark matter halo mass function, and \cite{Dutton2014} for the concentration-halo mass relation.
The results show a different choice of dark matter halo population modeling will not change the final event rate by orders of magnitude.

\begin{figure}[htbp] 
   \centering
   \includegraphics[width=\columnwidth]{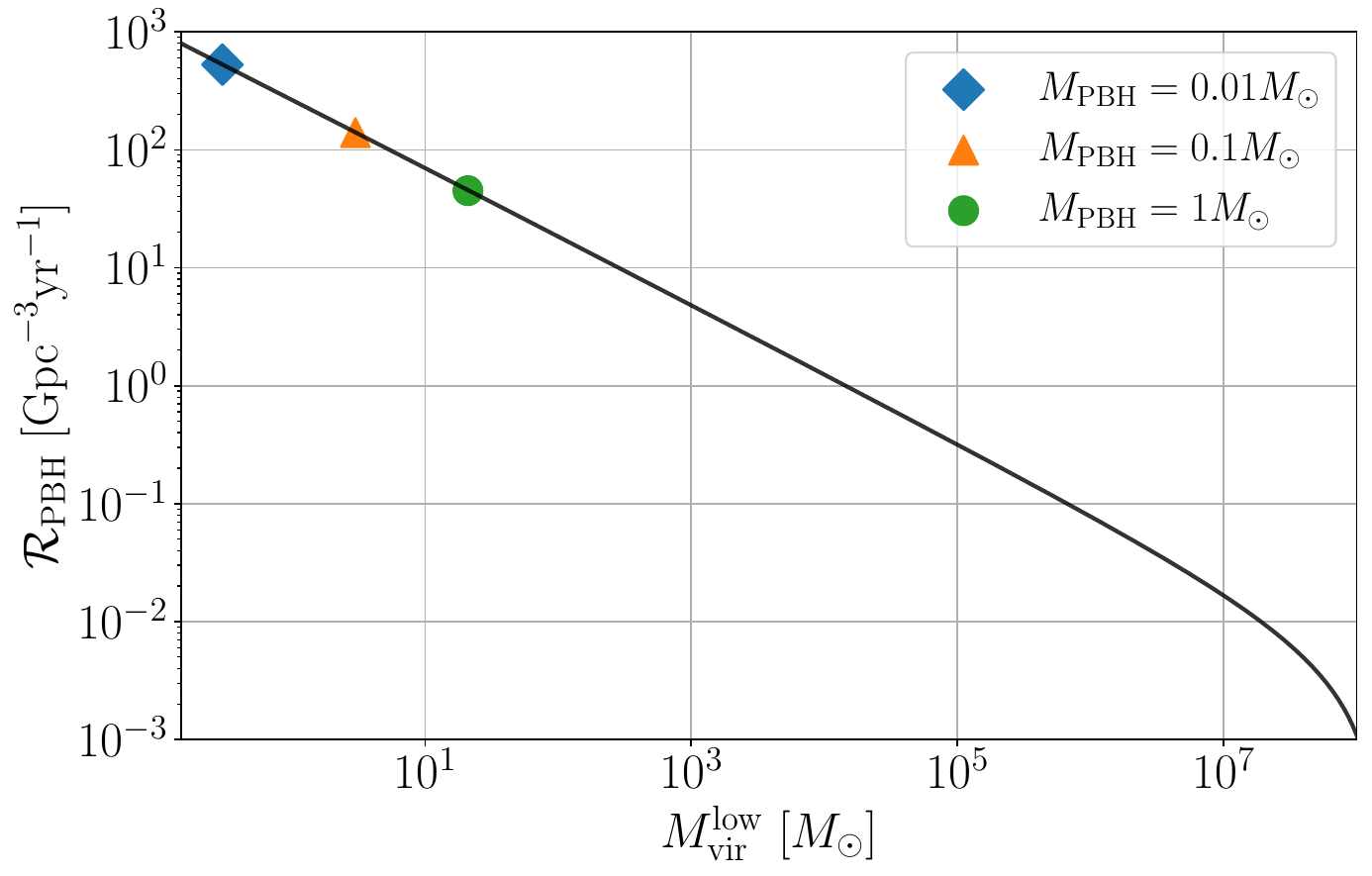} 
   \caption{Event rate for binary primordial black hole merger as a function of the lower cutoff of dark matter halo mass.
   The markers are the event rate for $M_\mathrm{PBH} = 0.01/0.1/1 \msun$, which are $528/140/45$ Gpc$^{-3}$ yr$^{-1}$, respectively.
   }
   \label{fig:rate}
\end{figure}

Assuming $f_\mathrm{PBH}=100\%$, the result of \cref{eq:rpbh} is shown in \cref{fig:rate}.
The main contribution to the integral in \cref{eq:rpbh} is from low-mass dark matter halos, therefore the result depends sensitively on the lower end for halo mass.
To illustrate the dependence on lower mass cutoff $M_\mathrm{vir}^\mathrm{low}$, we plot $\mathcal{R}_\mathrm{PBH}$ as a function of $M_\mathrm{vir}^\mathrm{low}$ in \cref{fig:rate}.
Also note that the integrand of \cref{eq:rpbh} does not depend on $M_\mathrm{PBH}$ due to compensation between the primordial black hole number density and the cross section.
To determine the lower limit of the dark matter halo mass for $M_\mathrm{PBH} = 0.01/0.1/1~\msun$, we follow the criterion by Eq. (11) of \cite{Bird:2016} requiring that the dark matter halo evaporation time scale due to dynamical relaxation exceeds the dark energy domination time scale ($\sim 3\times 10^9$ yr).
This choice results in the lower integral limit of \cref{eq:rpbh} being $0.3/3/21~\msun$ for $M_\mathrm{PBH} = 0.01/0.1/1~\msun$.
Choosing a lighter dark matter halo cutoff would contribute more binary primordial black hole merger events if the halos would not have evaporated at present and the NFW profile is still preserved.
High resolution dark matter N-body simulations are needed to precisely resolve the smallest structure of subsolar mass primordial black hole halos but this is beyond the scope of this paper.

With the above choices of lower integral cutoff, as shown in \cref{fig:rate}, the event rate for $M_\mathrm{PBH} = 0.01/0.1/1 \msun$ is $528/140/45$ Gpc$^{-3}$ yr$^{-1}$.
As a comparison, for $M_\mathrm{PBH} = 30 \msun$, the event rate is $\mathcal{O}(1)$ Gpc$^{-3}$ yr$^{-1}$ as computed in \cite{Bird:2016}.
Therefore, the merger rate of subsolar primordial black holes increases by one to two orders of magnitude compared to that of $30 \msun$.

To compare with observations, the sensitive volume,  which is defined to be the orientation-averaged volume assuming a match-filtering signal-to-noise ratio (S/N) $>8$ \citep{1304.0670}, is $\mathcal{O}(10^{-7})/\mathcal{O}(10^{-4})/\mathcal{O}(10^{-2})$ Gpc$^3$ for $M_\mathrm{PBH} = 0.01/0.1/1 \msun$ for Advanced LIGO at its design sensitivity.
Given the above estimated event rate, under our assumption of delta function mass distribution, the second generation detectors are capable of detecting $M_\mathrm{PBH} = 1 \msun$ sources with a $\mathcal{O}(1)$ year run, and would not be likely to observe $0.1 \msun$ and $0.01 \msun$ sources. 
Third generation detectors such as the Cosmic Explorer \citep{1907.04833} and Einstein Telescope \citep{10.1088/0264-9381/27/19/194002} are expected improve the sensitive volume by three orders of magnitude compared to the second generation, thus they would be able to detect a delta function mass of $M_\mathrm{PBH}=0.1\msun$ source every $\mathcal{O}(0.1)$ yr, and a $0.01\msun$ source with a $\mathcal{O}(10)$ yr observation time.
Conversely, a nondetection in the future would put constraints on the abundance of primordial black holes formed in this late-universe channel.

We also compare the binary coalescence rate to that derived from primordial black hole binary formation in the early universe. 
Using the derivation by \cite{sasaki:2016}, the merger rate is $\mathcal{O}(10^8)/\mathcal{O}(10^7)/\mathcal{O}(10^6)$ Gpc$^{-3}$ yr$^{-1}$ for $M_\mathrm{PBH} = 0.01/0.1/1~\msun$ assuming $f_\mathrm{PBH}=100\%$.
Therefore, an early-universe formation scenario dominates the event rate for subsolar mass compact binary coalescence, but this formation channel is not expected to yield eccentric binaries because they would have been circularized in the local universe. There are also model uncertainties under active investigation about what fraction of primordial black hole binaries would be disrupted after formation in the early universe \citep{Ali-Haimoud:2017rtz,Raidal:2018bbj,Vaskonen:2019jpv,Jedamzik:2020omx}. 
If a significant fraction are disrupted, the merger rate for the early-universe formation channel would be suppressed.
Recently, \cite{Boehm:2020jwd} also pointed out that the merger rate of early Universe formation channel is overestimated by reanalysis of the primordial black hole binary formation criterion.

Compact binaries formed through dynamical interaction may retain eccentricity within the ground-based gravitational-wave detector frequency band.
In the following we determine the eccentricity distribution for binary primordial black hole merger events following the references \cite{Leary:2009,Cholis:2016}.

Right after formation, the initial semi-major distance of a binary is given by 
\be\label{eq:a0}
a_0 = \frac{Gm_1m_2}{2|E_f|}
\ee
where $E_f = 1/2\mu v_\mathrm{rel}^2 - \delta E_\mathrm{GW}$.
Combining \cref{eq:rp,eq:a0} and inserting the relation $r_{p,0} = a_0 (1-e_0)$, one obtains an expression for the initial eccentricity at binary formation of
\be\label{eq:e0}
e_0 (m_1,m_2,v_\mathrm{rel},b) = \sqrt{ 1- \frac{2|E_f| b^2 v_\mathrm{rel}^2}{(m_1+m_2)m_1m_2}}
\ee

After formation of the binary, the semi-major distance $a$ and the eccentricity $e$ gradually decay due to gravitational-wave emission, the time evolution equation is given by \citep{Peters:1964}
\bn\label{eq:dadtdedt}
 \frac{da}{dt} &=& -\frac{64}{5}
\frac{G^3m_1m_2(m_1+m_2)}{c^5a^3(1-e^2)^{7/2}}
\(1+\frac{73}{24}e^2 + \frac{37}{96}e^4\),\nonumber \\ 
 \frac{de}{dt} &=& -\frac{304}{15}
\frac{G^3m_1m_2(m_1+m_2)e}{c^5a^4(1-e^2)^{5/2}}
\(1+\frac{121}{304}e^2 \).
\en
 
Combining Eq.~(\ref{eq:dadtdedt}) yields the expression for $da/de$ which can be integrated analytically, the result is
\be\label{eq:finale}
a(e) = a_0 \frac{\kappa(e)}{\kappa(e_0)}
\ee
where $\kappa(e)$ is a function of $e$
\be
\kappa(e) = \frac{e^{12/19}}{1-e^2}\( 1+ \frac{121}{304}e^2\)^{870/2299}.
\ee
Substituting \cref{eq:a0,eq:e0} and $a = \sqrt[3]{G(m_1+m_2)/\pi^2 f_\mathrm{GW}^2}$ from the Kepler's third law where the frequency $f_\mathrm{GW}=10$Hz, the nonlinear \cref{eq:finale} for $e^\mathrm{10Hz}$ can be solved by numerically finding the root.

\begin{figure}[htbp] 
   \centering
   \includegraphics[width=\columnwidth]{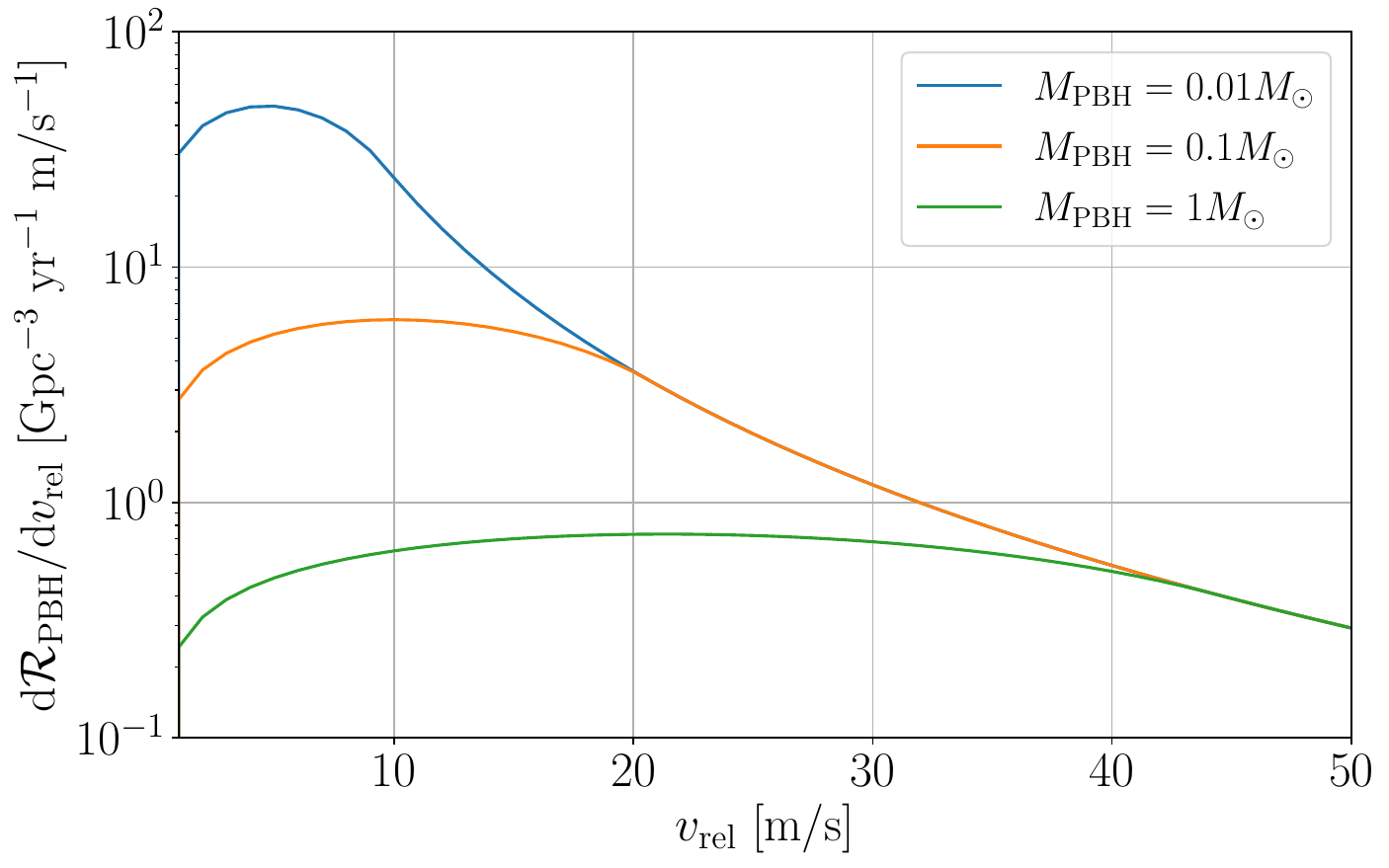} 
   \caption{The binary primordial black hole event rate density with respect to the relative velocity $v_\mathrm{rel}$ at the formation of primordial black hole binaries for $M_\mathrm{PBH}=0.01/0.1/1~\msun$.}
   \label{fig:vrel}
\end{figure}

As $e^\mathrm{10Hz}$ depends on $e_0$, which in turn depends on the initial relative velocity $v_\mathrm{rel}$ and the impact parameter $b$ for two black holes, we compute the relative velocity distribution by taking the derivative of $\mathcal{R}_\mathrm{PBH}$ with respect to $v_\mathrm{rel}$. 
The results are presented in \cref{fig:vrel}.
It can be seen that the initial relative velocities for $M_\mathrm{PBH}=0.01/0.1/1~\msun{}$ peak at $5/10/25$ m/s, respectively, which approximately correspond to the typical velocity of the lightest dark matter halo in the integral \cref{eq:rpbh}, because the merger events from the low-mass dark matter halo dominate the event rate.

We use a Monte Carlo method to simulate a population of binary primordial black holes with masses $0.01/0.1/1\msun$, respectively, each containing $\sim10^6$ sources, to obtain the distribution of $e^\mathrm{10Hz}$.
The square of the impact parameter, $b^2$, is chosen to be uniformly distributed in $[r_\mathrm{ISCO}^2, b_\mathrm{max}^2]$ \citep{Cholis:2016}, where $r_\mathrm{ISCO}$ is the innermost stable circular orbit for a Schwarzschild black hole (6 times the Schwarzschild radius), $b_\mathrm{max}$ is given by \cref{eq:cs}.
The initial relative velocity is drawn from the distribution in \cref{fig:vrel}.

\begin{figure}[htbp] 
   \centering
   \includegraphics[width=\columnwidth]{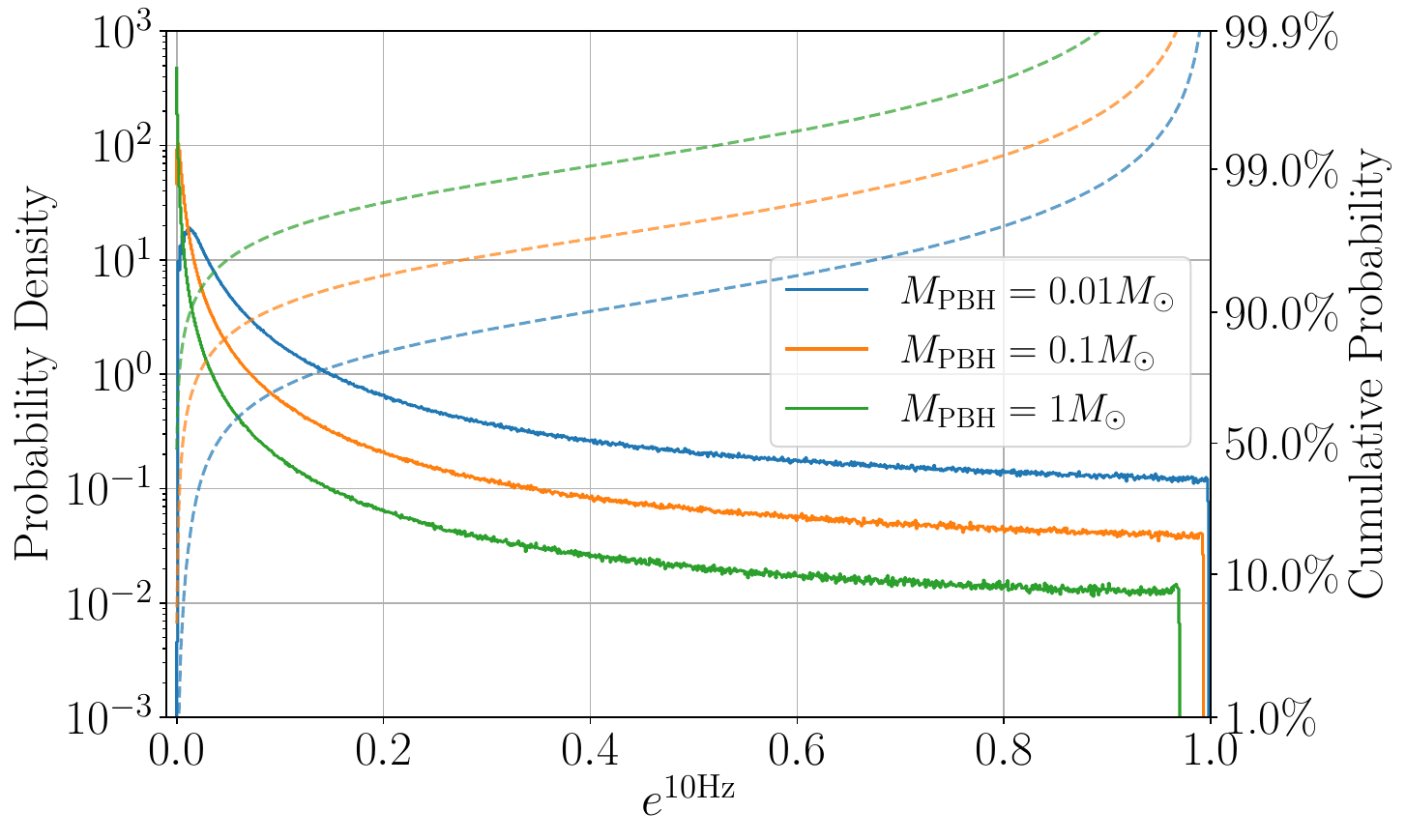} 
   \caption{The probability density distribution (solid lines, left y-axis) and the cumulative probability distribution (dashed lines, right y-axis) for the eccentricity of binary primordial black hole inspiral with component masses $0.01/0.1/1 \msun$ at gravitational-wave frequency $10$ Hz.} 
   \label{fig:ecc}
\end{figure}

The results for the distribution of $e^\mathrm{10Hz}$ are shown in \cref{fig:ecc} by the solid lines with the corresponding probability density distribution shown on the left vertical axis.
To assist understanding, we also plot the cumulative probability distribution in \cref{fig:ecc} as shown by the solid lines converging to $100\%$ at the right vertical axis.
As seen from the figure, lighter primordial black holes tend to retain larger eccentricity.
For $M_\mathrm{PBH}=0.01/0.1/1~\msun$, up to $89\%/40\%/12\%$ of the total inspiral events have $e^\mathrm{10Hz}\ge0.01$, and up to $29\%/9\%/3\%$ have $e^\mathrm{10Hz}\ge0.1$.
To connect these estimates with observations, assuming $f_\mathrm{PBH}=100\%$, we find that second generation detectors would be expected to detect an eccentric binary coalescence with $M_\mathrm{PBH}=1\msun$ within a few years observation time, while the third generations are expected to probe the eccentric events down to $0.01\msun$ with $10^3$ times of sensitive volume than Advanced LIGO and Virgo.
In addition, third generation detectors are designed increase sensitivity at frequencies of $\sim 2-5$ Hz, where the eccentricity will be more significant. Third generation detectors show promise for detecting or constraining our understanding of eccentric binaries.

\begin{figure}[htbp] 
   \centering
   \includegraphics[width=\columnwidth]{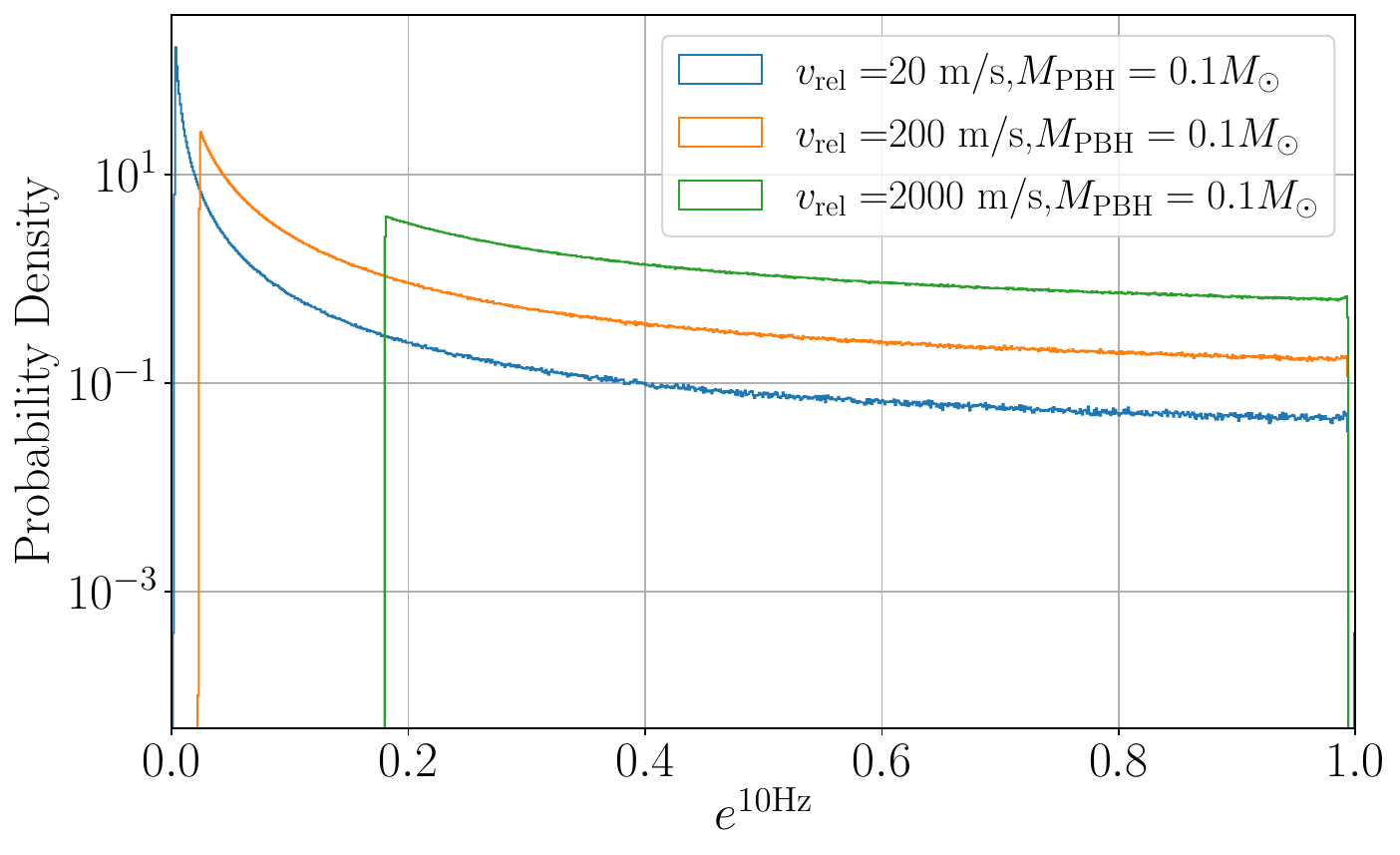} 
   \caption{The eccentricity distribution at 10 Hz for primordial black hole binaries with $M_\mathrm{PBH}=0.1~\msun$ for three choices of relative velocity, 20/200/2000 m/s.}
   \label{fig:com1}
\end{figure}

We further investigate the impact of $M_\textrm{PBH}$ and $v_\mathrm{rel}$ on the eccentricity distribution.
In \cref{fig:com1}, the mass $M_\textrm{PBH}$ is fixed to $0.1~\msun$, and the initial relative velocity $v_\mathrm{rel}$ is chosen to be $20/200/2000$ m/s.
The $e^\mathrm{10Hz}$ results show that a higher relative velocity leads to the formation of more eccentric binaries.
The cutoff for each distribution at the lowest eccentricity in \cref{fig:com1} represents the binaries that would have been circularized the most at the gravitational wave frequency $10$ Hz and are from those with the widest separation at initial formation and thus the lowest initial eccentricity, as can be seen from \cref{eq:e0}.

\begin{figure}[htbp] 
   \centering
\includegraphics[width=\columnwidth]{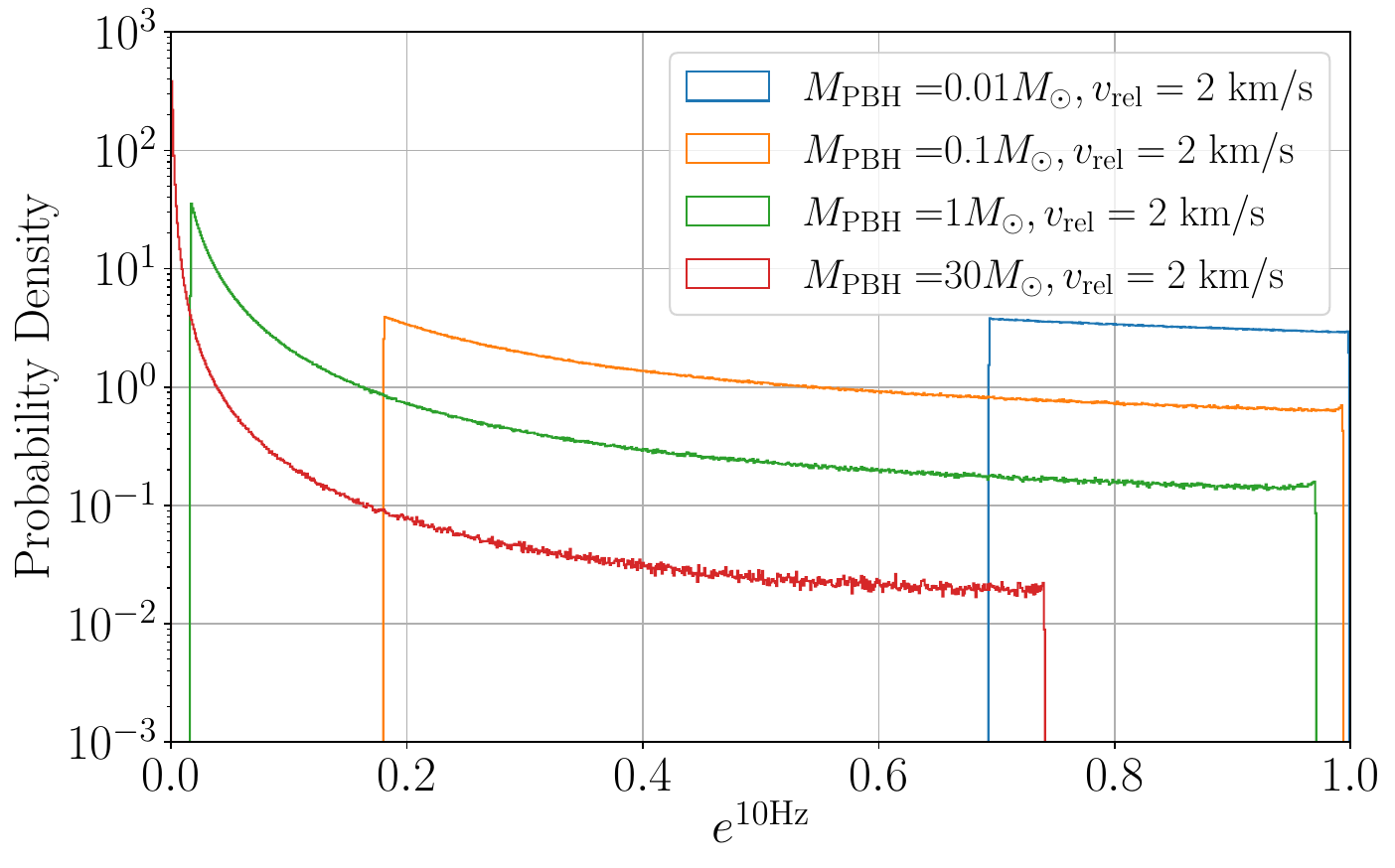} 
   \caption{The eccentricity distribution at 10 Hz for primordial black hole binaries with $M_\mathrm{PBH}=0.01/0.1/1/30 \msun$ with relative velocity fixed to 2 km/s.}
   \label{fig:com2}
\end{figure}

The eccentricity distribution for fixed $v_\textrm{rel}$ but different masses are shown in \cref{fig:com2}.
The relative velocity is chosen to be $2$ km/s, which is the typical velocity for $10^6~\msun$ dark matter halos.
The black hole mass is chosen to be $0.01/0.1/1/30~\msun$.
As considered by \cite{Cholis:2016} and shown in \cref{fig:com2}, a majority of $30~\msun$ binary black holes have eccentricity near $0$ at $10$ Hz.
However, the subsolar mass binaries all have nonzero peaks for eccentricity.
In particular, for $0.01~\msun$ primordial black holes in the $10^6~\msun$ dark matter halos, all binaries have high eccentricity not lower than $0.7$ at formation.
Similarly, the lowest eccentricity cutoff for each distribution represents the most circularized binaries in the simulated sources. 

The above numerical results show that subsolar mass compact binaries with high initial relative velocity tend to be more eccentric compared to the more massive binary black holes with low relative velocity.
Therefore, the impact of eccentricity on targeted gravitational-wave searches for subsolar mass compact binary coalescences deserves a more detailed study. 
We present our investigation in the next section.

\section{Impact of Eccentricity on Gravitational-wave Searches}\label{sec:search}

There has been one search targeting gravitational waves from eccentric compact binary coalescence using model-based matched filtering \citep{2020ApJ...890....1N}, where the authors searched for signals from eccentric binary neutron star mergers. The LIGO and Virgo Scientific Collaborations have also performed a nonmodelled generic search for eccentric binary black hole mergers \citep{Salemi:2019owp}.
Most targeted searches for compact binary coalescence use quasi-circular orbit waveform template banks, including the direct searches for subsolar mass black holes \citep{1808.04771,1904.08976,Nitz:2020}, 

As demonstrated in section \ref{sec:model}, subsolar mass primordial black holes can retain nonzero eccentricity in the LIGO and Virgo frequency band if they form in the late universe through dynamical capture.
We aim to quantitatively study the potential loss in detection sensitivity when using only quasi-circular waveform template banks to search for signals from eccentric binary primordial black holes. 
To quantify the S/N of gravitational-wave detection, we define the following noise-weighted inner product
\be
(h_1,h_2) = 4\Re \int \frac{h_1(f) h^\ast_2(f)}{S_n(f)}
\ee
where $h_1$ and $h_2$ are signals or gravitational-wave templates in the Fourier domain and $S_n(f)$ is the one-sided noise power spectral density.
In this section we only consider the Advanced LIGO design sensitivity in the frequency range of $[20,1024]$ Hz.
The matched-filter S/N between the detector output $s$ and the template $h$ is defined to be
\be
\rho = \frac{(s,h)}{\sqrt{(h,h)}}.
\ee

To measure the similarity between two waveform templates $h_1$ and $h_2$, the overlap function is defined as the normalized inner product
\be
\mathcal{O}(h_1,h_2) = \frac{(h_1,h_2)}{\sqrt{(h_1,h_1)(h_2,h_2)}}.
\ee
Two templates may be different up to a constant time and phase in the Fourier domain, thus the match function between two waveforms is given by maximizing over the offset of coalescence time $t_c$ and an overall phase $\phi_c$
\be\label{eq:match}
\mathcal{M}(h_1,h_2) = \max_{t_c,\phi_c}\left(\mathcal{O}(h_1,h_2e^{i(2\pi f t_c -\phi_c)})\right).
\ee

There have been several waveform approximants modeling eccentric compact binary coalescence, such as \cite{Moore:2016, Huerta:2018,1910.00784, 2001.11736}.
In this work, we utilize the model proposed by \cite{Moore:2019a,Moore:2019b,Moore:2018}, referred to as \texttt{TaylorF2e} \footnote{The source code can be found in https://github.com/gwastro/TaylorF2e}, for speedy waveform generation.
\texttt{TaylorF2e} is an inspiral waveform developed in the Fourier domain and is accurate to the third post-Newtonian order.
The eccentricity is valid up to at least $0.8$ in the low-mass case $\sim1\msun$ \citep{Moore:2019b}, which is sufficient for our purpose to investigate the eccentricity of subsolar mass compact binaries.

In order to extract gravitational-wave signals in the data, a targeted search \citep{Usman:2016,Messick:2017} builds a precalculated template bank to match filter the data.
To measure the detection ability of a template bank built with quasi-circular waveforms to search for eccentric binaries, we first generate a template bank by a geometric method based on hexagonal lattice placement \citep{Cokelaer:2007,Brown:2012qf} using the quasi-circular waveform approximant \texttt{TaylorF2} \citep{Buonanno:2009}. For this study, we assume the component black holes will have negligible spin.
The template bank is designed to recover sources with component masses in the range $[0.1,1]~\msun$.
The bank is discrete in the parameter space, and we require the maximum mismatch due to discreteness is no higher than $3\%$ by construction. The above setting results in a bank with $6995517$ templates. 

Next we generate simulated gravitational-wave data with a source population using the \texttt{TaylorF2e} approximant. 
The mock data are then compared with every template in the bank using \cref{eq:match} to obtain the maximum match ``fitting factor" over the entire template bank. A fitting factor of $x$ would lead to a $1-x^3$ loss of the detection rate for a search with a fixed amount of noise, as the gravitational-wave search S/N would be decreased to $x$ times the theoretically optimal value.

\begin{figure*}[htbp] 
   \centering
\includegraphics[width=\columnwidth]{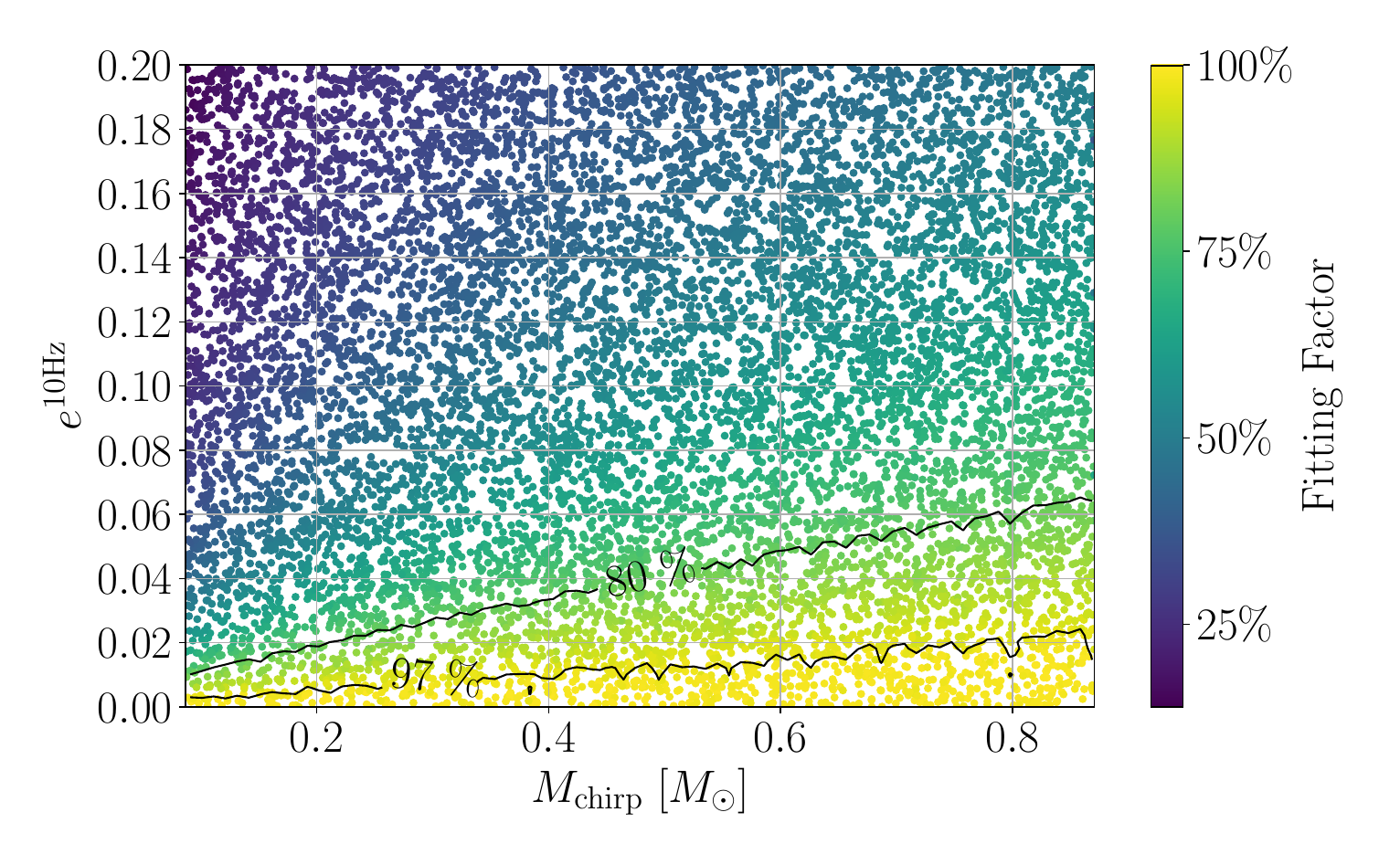} 
\includegraphics[width=\columnwidth]{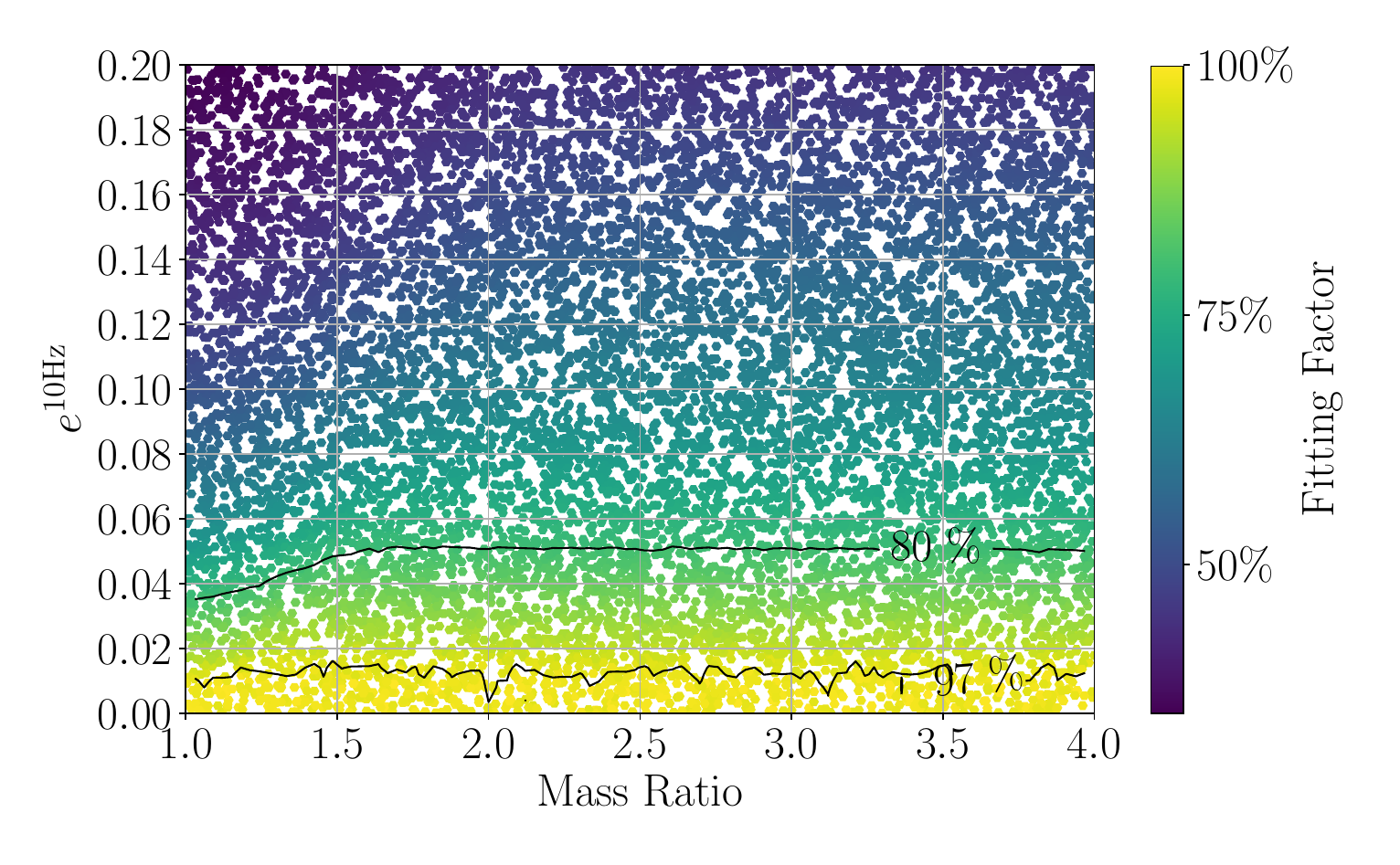} 
\caption{The fitting factor for simulated data of eccentric subsolar binary black hole coalescences with respect to the template bank constructed from a quasi-circular waveform approximant. 
Each point represents an injected mock signal.
In the left panel, the signals are generated from equal-mass binaries with a component mass in $[0.1.1]~\msun$ uniformly.
In the right panel, the chirp mass is fixed to $0.4~\msun$ and the mass ratio is in $[1,4]$.}
   \label{fig:match-equalmass}
\end{figure*}

In \cref{fig:match-equalmass}, we generate  eccentric subsolar mass binary black hole gravitational-wave signals and compute the fitting factor with respect to the noneccentric template bank.
Each point in the figure represents a mock signal injection.

In the left panel of \cref{fig:match-equalmass}, the binaries are chosen to have equal mass. $M_\textrm{PBH}$ and $e^\mathrm{10Hz}$ are uniformly distributed in $[0.1,1]~\msun$ and $[0,0.2]$, respectively.
The fitting factor for $e^\mathrm{10Hz}$ and the chirp mass $M_\mathrm{chirp}=(m_1m_2)^{3/5}(m_1+m_2)^{1/5}$ ($\sim 0.87M_\mathrm{PBH}$ for equal mass) is presented.
The figure also illustrates the $97\%$ and $80\%$ contour lines obtained from fitting the plot.
The figure shows the trend that the fitting factor gets worse for higher eccentricity, and also for lighter binaries because of their longer duration. For a component mass $m=0.1/1 \msun$, the fitting factor drops below $97\%$ with $e^\mathrm{10Hz}\ge0.003/0.20$.

In the right panel of \cref{fig:match-equalmass}, we consider the impact on fitting factor from different mass ratios by fixing $M_\textrm{chirp}=0.4~\msun$. 
The mass ratio is uniformly distributed in $[1,4]$ and the eccentricity is uniformly in $[0,0.2]$. 
As shown, the $97\%$ and $80\%$ fitting factor lines only depend weakly on the mass ratio and decrease for higher mass ratio.

\begin{figure}[htbp] 
   \centering
\includegraphics[width=\columnwidth]{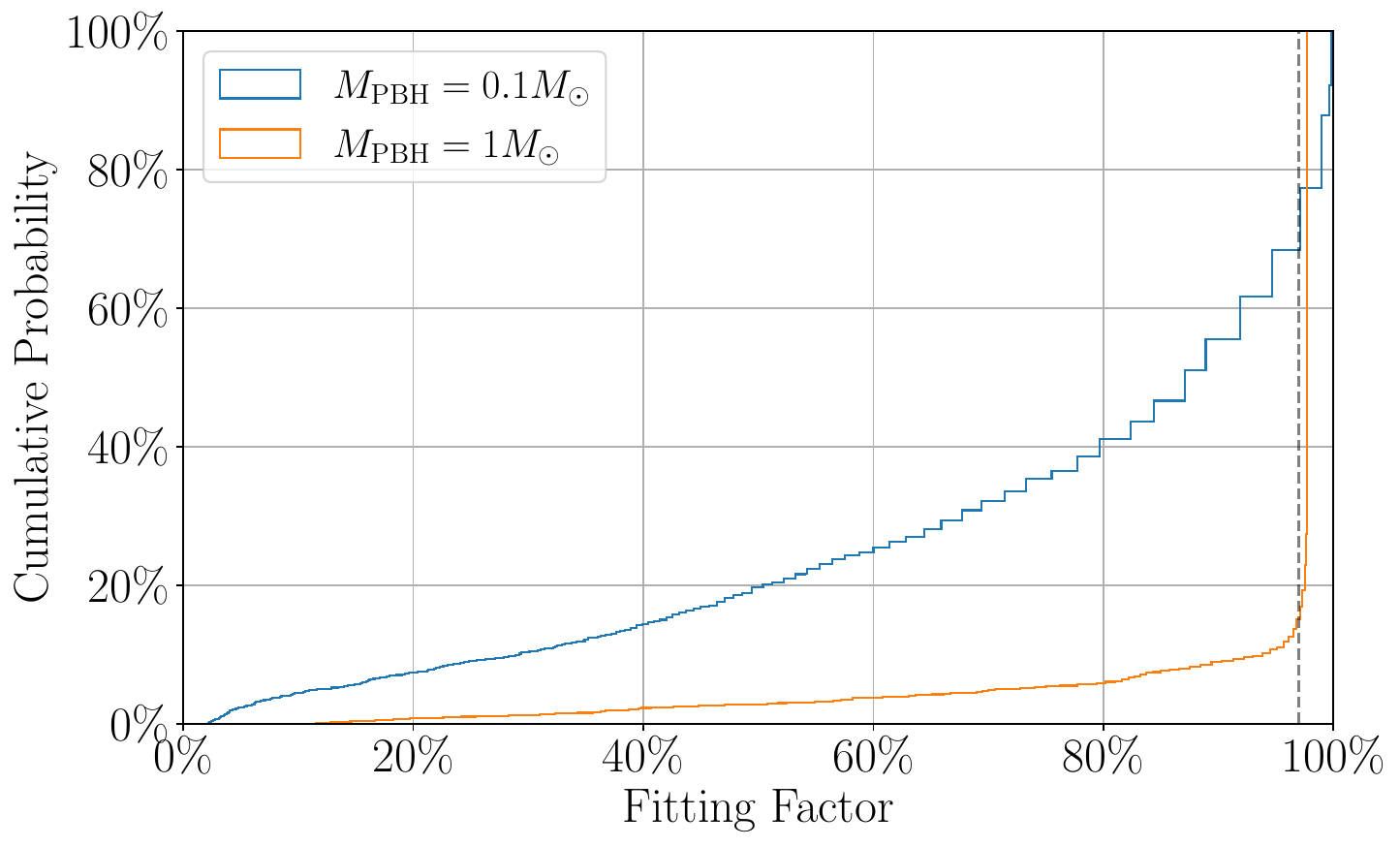} 
   \caption{Cumulative distribution for the fitting factor for gravitational-wave sources with $0.1\msun$ and $1\msun$ and eccentricity distributions derived from Section \ref{sec:model}.
   The vertical dashed line denotes $97\%$, which represents the lower limit for the fitting factor for signals without eccentricity.}
   \label{fig:real}
\end{figure}

We also consider more realistic distributions for $e^\mathrm{10Hz}$ as derived in \cref{fig:ecc} of Section \ref{sec:model}.
We generate two groups of mock data each with $\sim2000$ signals using \texttt{TaylorF2e}.
The component mass is $0.1\msun$ and $1\msun$, respectively, and only equal-mass binaries are considered.
The inclination angle is chosen to be uniformly distributed in $[-\pi/2,\pi/2]$, the polarization angle is uniformly in $[0,2\pi]$, and the source sky location is isotropic.
The eccentricity distribution is drawn from \cref{fig:ecc} for $0.1\msun$ and $1\msun$ sources accordingly.

The cumulative probability with respect to the fitting factor is shown in \cref{fig:real}.
The vertical dashed line in \cref{fig:real} denotes $97\%$, which all the signals with zero eccentricity should exceed by construction of the template bank.
However, due to the existence of eccentricity, for $M_\textrm{PBH}=0.1\msun$, up to $68\%$ of the signals have a fitting factor lower than $97\%$, and $41\%$ have a fitting factor lower than $80\%$.
For $M_\textrm{PBH}=1~\msun$, $15\%$ of the signals have a fitting factor lower than $97\%$, and $6\%$ of the signals have a fitting factor lower than $80\%$.
Also note that, when the fitting factor is low, the mismatch between signals and templates may be significant enough to cause issues with signal-based vetoes~\citep{Allen:2004gu} which would further reduce the detection rate of eccentric sources.
Therefore our results based on the fitting factor are a conservative estimation for the loss of detection rate.

Given that the fitting factor can only measure the fractional recovered S/N for a single injected signal, we use the effective fitting factor ${FF}_\mathrm{eff}$ \citep{2016PhRvD..94b4012H,2014PhRvD..89b4010H,Buonanno:2002fy} to account for the overall fraction for detection loss for a whole group of simulated signals.
The effective fitting factor  is defined as a mean average of fitting factors weighted by the signal amplitude
\begin{equation}
    {FF}_\mathrm{eff} = \left(\frac{\sum_{i=1}^{N} {FF}_i^3 \sigma_i^3  }{\sum_{i=1}^{N}  \sigma_i^3 }\right)^{1/3},
\end{equation}
where the subscript $i$ denotes the i-th mock injection, $N$ is the total number of signals, ${FF}_i$ is the fitting factor, and $\sigma_i = \sqrt{(h_i|h_i)}$ is the optimal S/N for the i-th signal. An effective fitting factor of $x$ would lead to a $1-x^3$ loss of detection rate.

For our two groups of injection with $M_\textrm{PBH}=0.1~\msun$ and $1\msun$, ${FF}_\mathrm{eff}=83\%$ and $95.8\%$, respectively, which corresponds to an overall loss of $42\%$ and $12\%$ of signals compared to the idealized maximum.
For comparison, if only selecting the injected signals with $e^\mathrm{10Hz}=0$, the effective fitting factor is $99.7\%$ for $M_\mathrm{PBH}=0.1\msun$ and $97.8\%$ for $M_\mathrm{PBH}=1\msun$, which corresponds to a $0.9\%$ and $6.4\%$ loss, which arises from the discreteness of template bank.
After subtracting the signals missed due to template bank discreteness, for $M_\textrm{PBH}=0.1\msun$ and $M_\textrm{PBH}=1~\msun$ sources within the late-universe dynamical encounter model, we conclude that the eccentricity effects can lead to loss of $41\%$ and $6\%$ of the signals, respectively, by a targeted search with a quasi-circular waveform template bank using Advanced LIGO at its designed sensitivity.
For primordial black holes with an extended mass in the range $[0.1,1]\msun$, we expect the the corresponding results for loss of detection should lie in between these.

\section{Discussions and conclusions}\label{sec:con}

In this work, we investigate the prospects for detecting gravitational-wave signals from eccentric subsolar mass binary black hole inspirals, which would be a signature for primordial black hole binaries formed by dynamical capture in the local universe.
The merger rate and eccentricity distribution for binaries assuming a delta function mass distribution varying in $[0.01,1]\msun$ are derived.
For $1\msun-1~\msun$ binaries, the detection rate is $\mathcal{O}(1)$/yr for Advanced LIGO and Virgo with design sensitivity, if primordial black holes account for a majority of dark matter, and $12\%(3\%)$ of the sources have $e^\mathrm{10Hz}\ge0.01(0.1)$, thus eccentric signals may be detected with a few years of observation based on the assumed model.
For $0.1-0.1\msun$ and $0.01-0.01~\msun$ binaries the detection rate would be too low for the second generation ground-based detectors, but it is promising for detection by the third generation detectors.
Conversely, nondetection in the future would put constraints on our understanding of this formation channel.

We also investigate what fraction of the eccentric binaries can be missed by using quasi-circular waveform template bank.
Results show that current targeted searches can miss  $41\%$ and $6\%$ of the events with component masses of $0.1\msun$ and $1\msun$, respectively, with Advanced LIGO designed sensitivity, due to the eccentricity of primordial black hole binaries arising from dynamical encounter in the local universe.

For the formation scenario of primordial black hole binaries, we only consider the two-body direct capture through gravitational-wave braking.
Other mechanisms such as three-body interaction via Kozai-Lidov effects \citep{Kozai,Lidov} may further improve the event rate, and would be interesting to consider in a future work.

Third generation gravitational-wave detectors such as Cosmic Explorer \citep{1907.04833} and Einstein Telescope \citep{10.1088/0264-9381/27/19/194002} are being planned.
The detection volume is expected to increase by three orders of magnitude compared to the second generation.
Moreover, the sensitive band can be pushed down to $\sim 2$ Hz, at which the eccentricity for primordial black hole binaries is more significant than later at 10 Hz as considered in this work. Third generation detectors are even more promising for hunting for eccentric gravitational-waves signals.

The code and data associated with this work are available at \url{https://github.com/gwastro/prospects-subsolarmass-ecc}.

\acknowledgments
We acknowledge the Max Planck Gesellschaft for support and the Atlas cluster computing team at AEI Hannover.

\bibliographystyle{aasjournal}
\bibliography{ads.bib}
\end{CJK*}
\end{document}